# The Discovery of the First Millisecond Pulsar: Personal Recollections

S. R. Kulkarni

Owens Valley Radio Observatory 249-17, California Institute of Technology,
1216 E. California Blvd., Pasadena, CA 91125, USA

**ABSTRACT**

This article provides a first-hand account of the 1982 Arecibo observations that led to the discovery of PSR B1937+21, the first-known millisecond pulsar. It is a companion paper to Demorest & Goss (2024) and Readhead (2024).

## 1  Prologue

I joined the graduate program of the department of astronomy, University of California at Berkeley (UCB), in the Fall of 1978. My goal was to pursue a thesis in experimental radio astronomy. The Radio Astronomy Laboratory (RAL) of UCB was very active with projects in millimeter astronomy (Jack Welch), atomic hydrogen (Carl Heiles) and VLBI (Donald Backer). During my first year I earned my keep as a technician of RAL, specifically wire-wrapping control boards and soldering boards for a 3-level correlator that was being built for UCB's 85-foot radio telescope located at the Hat Creek Radio Observatory. In the summer of 1979, I attended the National Radio Astronomy Observatory (NRAO) summer program held at Charlottesville. I worked with Dr. Craig Walker and analyzed Very Long Baseline Interferometry (VLBI) data on W3OH. I took the occasion to read up all the memos and the manual for NRAO's Mark II VLBI correlator. Many memos were written by Barry G. Clark, and I was in awe of his talent. In those days, VLBI was done in only one polarization (to minimize cost). I felt I had mastered VLBI and so upon return to Berkeley I began a pilot project on "dual polarization" VLBI under the supervision of Backer. I worked on hardware (electronics for recording a second channel) and visited Caltech to scope out the programming changes needed to process dual-polarization signals at Caltech's "network" correlator. By the end of the year, I realized that perhaps this project was ahead of its time and that I should choose a project that could be completed on the timescale of a thesis.

Post-doctoral fellow Dr. Leo Blitz got me working on the distribution of HI in the outer Galaxy, a relief from soldering and wire-wrapping. This eventually led to two papers. In those days, foreign students had an additional advisor ("mentor"), perhaps to ease the transition from their culture to the US culture. By an amazing stroke of luck, Professor Chris McKee was assigned to be my mentor. Thanks to McKee and Professor Charles Townes (who had started an experimental program of far-IR fine structure line studies of the Galactic interstellar medium; I audited his class) I got very interested in the interstellar medium (ISM). The three-phase McKee-Ostriker model published in 1977 was making big waves. Discussions with Heiles led to a new thesis project: probing the temperature distribution of Galactic HI via 21-cm absorption spectroscopy of bright radio sources. This project had a great radio astronomy tradition: the Caltech thesis project of Barry Clark using the Owens Valley Radio Observatory (OVRO) spectral-line interferometer (Clark, 1964) and foundational work at Parkes Observatory (Radhakrishnan & Murray, 1969). These two efforts motivated the two-phase model of the Galactic ISM (Field et al., 1969).

The plan was to build an air-linked interferometer between the 1000-foot Arecibo radio telescope located in Puerto Rico and a 30-foot telescope located at Los Canos, about 10 km NNE of the Arecibo dish. The latter was originally built by NASA to help resolve East-West ambiguity in S-band radar Doppler reflectometry of rotating planets  The project consisted of (1) instrumenting the Los Canos dish with an L-band feed and a cooled L-band receiver, (2) building a radio-link for the local oscillator and (3) correlating the data in real time. I began an apprenticeship program of building RF amplifiers under the supervision of RAL staff, W. T. (Tap) Lum and Dr. David R. W. Williams. Professor Anthony (Tony) C. S. Readhead, Director of Caltech's Owens Valley Radio Observatory (OVRO), promised to provide us with an L-band feed. Finally, thanks to my Charlottesville stint and detailed understanding of the Mark II VLBI correlator I came up with a plan to convert the Arecibo digital auto-correlator into a cross-correlator complete with lobe rotation, phase switching and delay compensation.

## 2    Very Large Array

By Fall of 1980 I was ready to head to Arecibo. Unfortunately, there was a delay in the delivery of the L-band feed from OVRO. Fortunately, RAL post-doctoral fellow Dr. Baudwijn Baud who had obtained some time at the newly-commissioned VLA was seeking a research assistant to accompany him to Socorro and reduce OH maser data. Being at loose ends I tagged along. The data reduction was slow with a long interval between job submission to the queue of Digital Equipment's DEC-10 main-frame computer and the results coming out. The long intervals gave me an opportunity to systematically read and digest the VLA memos. I came up with a new mode for the VLA: phasing the three arms of the VLA followed by correlating the summed outputs. The resulting 3-element interferometer could be used to obtain HI absorption spectra of point sources with minimal need for computing resources.

Dr. Ron Ekers, the site Director, not only provided the personnel to prove the proposed method but after my successful demonstration allocated large amounts of unused time (i.e., between 6 pm and 6 am) to my project. However, lodging at the Visiting Scientists Quarter (VSQ) was tight and given my anticipated long stay (several months) I was asked to move to a construction trailer at the site (provided at no cost, to the relief of Heiles!). Living in a trailer – with loud clanks as the Sun rose and set – brought a concrete awareness of the different coefficients of expansion of aluminum and iron. I spent the hot summer of 1981 working in the fashion of owls, awake at night and asleep during the day. These observations led to my first thesis paper (Dickey et al. 1983).

The VLA being brand new (formally inaugurated in October 1980) meant that the crème de la crème of the radio astronomy world was coming to the VLA to obtain and analyze data from this marvelous new facility. The stay at the VLA site expanded my pool of radio astronomy friends and colleagues. I also started taking interest in interferometric imaging and was particularly impressed with the power of a new computing device which was attached to the main-frame computer – Floating Point System's Array Processor 120B (AP-120B). Unlike computers, an array processor, as implied by its name processed not scalars but vectors. Thus, two 256-element vectors can be multiplied in one cycle – an amazing speed-up of a factor of 256 over a computer. The AP-120B device had a reputation of being very difficult to program, though. It was generally rumored that only two people at the VLA (B. Clark and an ace programmer) knew how to program the AP-120B.

## 3    Arecibo

In the Fall of 1981, the OVRO L-band feed arrived, and it was shipped off to Arecibo Observatory which was run by National Astronomy & Ionospheric Observatory (NAIC) and managed by Cornell University. In the winter of 1981 I headed to Arecibo. Before I left Berkeley, I expressed a strong interest to Backer to work on a pulsar project during my stay at Arecibo. He mentioned an object, 4C21.53 which he thought was a young pulsar like the Crab pulsar. However, his past efforts had not yielded pulsations.

Puerto Rico was exotic. The language (Spanish) was new and the food very different. The cook at the Observatory's cafeteria had a hard time understanding the concept of vegetarianism. However, there were compensations. I got to use the world's largest radio telescope, I became a fan of salsa music, made several local friends and attended weekend salsa dance parties at the local Pentecostal church.

Astronomers who were allocated observing time had to conduct the observations in person. Arecibo is a transit telescope. A given object is accessible to observations for, at most, three hours. A group of astronomers interested in the study of, say, the Pisces-Perseus super-cluster, would need the telescope for a certain range in LST (Local Sidereal Time) while those interested in the Galactic plane another range in LST and so on. The backend set up was different for each methodology (continuum, HI, pulsars, planetary radar, ionospheric studies). The stay at Arecibo, like that at the VLA, increased my circle of radio astronomy friends.

I had to work on many fronts, and it became clear that I would have to stay at Arecibo for quite some time, possibly six months or longer. Heiles did not have the funds to support me for such a long period (VSQ lodging). The Arecibo site director, Dr. Donald C. Campbell, kindly agreed to provide lodging and boarding and a car with the proviso that I work part time as "friend of the telescope". This meant re-cabling the radio frequency (RF) sub-system, the intermediate frequency (IF) sub-system and arranging the appropriate digital backend as the users rotated off. The ionospheric and planetary radar groups had their own crew. Nonetheless, I used the occasion to learn the tools of the trade in those two fields.[1] Most astronomy users needed help. I noticed that most of the pulsar groups used their own bespoke setups and so required no help from me. [The pulsar field was very competitive those days. As an instance of that, I give one example: only much later I came to know that Val Boriakoff, a staff member at Arecibo, had conducted search for pulsations from 4C21.53 earlier in March of 1982.]

I was very happy with the part-time job since it played right into my internal and secret view of what I desired of my thesis – namely to become a master of all techniques of observational radio astronomy. Equipped thus I could pursue questions in astronomy that could only be addressed by an experienced observer. I refer to my aspiration as a secret (in the sense that I did not advertise this view) because the classical approach is to first ask the question and then acquire the essential background and skills to answer that question. I had two local mentors: Dr. Kenneth Turner who was my official supervisor for the interferometer project and Dr. Michael Davis, the head of the Radio Astronomy group and who had wide knowledge of observing strategies, algorithms and computing. I was a frequent weekend guest at their beach-side houses located Ramey ex-Air Force Base (Aguadilla).

A typical 24-hour period at Arecibo had half a dozen observational programs. There were usually gaps between programs to allow for switchover and for maintenance (including pointing runs). A great advantage of my job was that I got the first dibs at any left-over time. Overall, I probably had about twenty hours of observing time through this scheme! I used this time for pedagogy (e.g., measuring the brightness temperature of the moon, measuring the oscillation of the azimuth arm), undertaking science (measuring rotation measure of bright extra-galactic sources) and even some fun – working with engineer Robert (Zimmo) Zimmerman who was also a ham operator, moon bouncing of ham radio signals! These forays made me quite familiar with the continuum backend, namely square law detectors and the data acquisition (D/A) system ("ADAGIO"; named by engineer Peter Shamus (now at the Jet Propulsion Laboratory), classical music afficionado, and justified because the analog to digital to data system (A/D) was supplied by ADAGE Inc and the "IO" stood for Input/Output. The irony, as will become clear from the discussion below, is that *adagio*, in western classical music, stands for slow tempo!). I found planetary radar not only fascinating but rich in digital coding.

Over time, at Arecibo, thanks to my part time job, I became conversant with many branches of astronomy: continuum, HI (both Galactic and extragalactic) and acquired passing familiarity with ionospheric studies and planetary radar but not pulsar astronomy. By late August, my interferometer started working (Kulkarni et al., 1985). I submitted an ambitious proposal seeking significant observing time for my thesis project. With the commissioning of the interferometer, I thought I would spend a week on the pulsar project that Backer and I discussed before I returned to Berkeley. I only allocated a week because at this point, I had been cooped up at the Observatory for over five months (or so). I was suffering from severe cabin fever made only worse by the fact that I was missing all my friends in Berkeley.

I reviewed the notes that Backer had given me before I left for Berkeley to Arecibo (late winter of 1981 or January of 1982). At that time Backer's conjecture was that 4C21.53 consisted of a steep spectrum source and a flat spectrum nebula, in the manner of the Crab Nebula and Crab pulsar (with the pulsar being the steep spectrum source and the plerion being the flat spectrum source). Since the birth period of the Crab pulsar was estimated to be about 16 milliseconds the plan was to sample the detected signal at 4 milliseconds. I approached two pulsar hunters to get some tips of pulsar searching but did not get much help. Fortunately, Arecibo Observatory had a good library, with a complete collection of Arecibo based theses. I found the thesis of J. M. Comella and papers by Richard Lovelace to be particularly useful for a pulsar tyro. These papers introduced me to harmonic summing, stacking of power spectrum and fast folding algorithm.

I had developed enough understanding of the hardware but soon realized that Backer had not addressed the issue of analysis at all. An essential step in pulsar analysis is Fourier transforming the sampled time series. The sensitivity of a pulsar search scales with the square root of the duration of the observation. The input time series is broken into many such chunks and the power spectrum added to increase the sensitivity. The Observatory's Harris/6 computer was, even by the standards of those days, a mini-computer and so had limited "core" memory. I spent a few days and developed a disk-based Fast Fourier Transform (FFT) routine. I do not recall the maximum length of the transform (likely 32K points). The routine was exceedingly slow.

An AP-120B was attached to the Observatory's Harris computer. I decided that I had no choice but to invest time and learn how to program the AP120B. The VLA rumor mill was indeed correct. The AP-120B was a beast to program. After many days I was able to bend the AP-120B to my will and undertake relatively long transforms (64K or 128K, I cannot recall) quite rapidly.

Finally, I had identified an hour slot of open time centered on 19 hours LST on 25 September 1982 which was well suited to observe 4C21.53. I requested Davis to help me set up the observations. The standard ADAGIO continuum set up sampled the signal at a tenth of a second (10 Hz). Backer's plan called for sampling at 250 Hz. The anti-aliasing filter that was needed for this sampling rate used op-amps. I did not have the time to verify the filter worked as labeled. Rather than risk using something that I had no experience with I decided to make my own passive R-C filter, but the time constant was small, about a millisecond. My plan was to decimate the data to the required sampling rate, post-detection. I thought this set up maximized the phase space of observations. The choice of the passive simple R-C filter was a lucky break. Had I stuck to the planned 4-millisecond sampling time and used the high-quality (sharp edge) anti-aliasing filter I would have missed the detection of millisecond pulsations. As will become clear from the discussion below the poorer quality of my anti-aliasing filter turned out to be helpful.

The observations began at dusk. The plan was straightforward – observe for ten minutes on the source and a few minutes off source and repeat. The data was recorded via ADAGIO. The data recording speed was set to match the sampling frequency. I was not aware that the recording speed was higher than the maximum allowed rate. This led to a red herring, discussed below.

The observations concluded an hour later, and Davis left for Ramey. I had a quick supper and began analysis of the data. By mid-night I had discovered two peaks, a weak fundamental and a strong harmonic in the first on-beam dataset. The adjacent off-beam data was devoid of the peaks. It was an exciting moment, and I had trouble sleeping. However, the signal was not seen in the second on-beam dataset. The next day after I woke up and after breakfast, I informed everyone that my AP-120B program had worked and I had found a millisecond pulsar (in that order). I then waited until it was morning in California and called up Backer. I informed him that 4C21.53 was a millisecond pulsar. Backer went silent and then said, "Do not mention this to anyone." Much to Backer's chagrin, I replied by saying that I had advertised to all and sundry that thanks to my AP-120B program I had found a millisecond pulsar.

On Monday, the next day, the Observatory was abuzz with my discovery. Campbell asked me to write a short proposal so that he could assign a proper "Project number" to the proposal (as opposed to charging to Project X which was the catch-all for calibration and miscellaneous usage). Campbell wanted me to stay a week and confirm the results and understand why the signal was not detected in the second data set. I was, as mentioned earlier, suffering from cabin fever and sorely needed a break. Furthermore, I had confidence in my set up, analysis and thus the results. I thought the source was variable and confirmation could wait until my return to Arecibo. Despite Campbell's pleas I flew out of San Juan to San Francisco two days later.

## 4      Return to Arecibo

Backer, Heiles and I returned in November. We had a two-pronged plan: I was responsible for the pulsar search while Backer and Heiles would look for scintillations which is a hallmark of super compact objects (and, at that point, the only known objects to have exhibited deep scintillation at GHz frequencies were pulsars). Both observations were undertaken on our first evening. For the pulse search Davis and I used the radar A/D system since this system, unlike ADAGIO, was explicitly designed to handle high sampling rates. For the signature of scintillations Heiles and Backer used the Observatory correlator with modest integration time (minutes). I had my analysis pipeline ready and so was able to analyze the data after the observations ended. Heiles worked through the night and developed a program to display the two-dimensional dynamic spectra.

That night I confirmed that the source was still pulsing. However, the period was 1.557,807 milliseconds whereas the period I found in September was 1.557,708 milliseconds. The inferred period derivative, *dP/dt* of $3 \times 10^{-14}$ s/s implied a slowing-down timescale of only a thousand years! Next day, Heiles' line printer plot of the dynamic spectrum showed signs of deep scintillation (in L-band). The deep scintillations explained both the strong detection and non-detections of my September pulse-search observations. With the situation clarified, Backer decided to send out an IAU Circular announcing the discovery of the pulsar Backer et al. (1982a). The position of the source (from VLA and WSRT imaging of this field provided by Miller Goss), the discovery period, the apparent *dP/dt* and a rough dispersion measure were included in this Circular.

The short slowing down timescale was a big topic for discussion. I must admit that as a young graduate student I was unconcerned mainly because I could not grasp the big issues that were at stake. Apparently, we were either extra-ordinarily lucky to observe an effervescent source or that there existed a very large population of millisecond pulsars or something else! There were serious discussions whether the source was slowing down by emission of gravitational waves.

The few observations in November showed that *dP/dt* < $10^{-15}$ s/s. During November, most of my effort went into analyzing possible problems with the data sampling set up Davis and I had devised for our September observations. After considerable amount of sleuthing, I realized that I had run ADAGIO at a sampling rate larger than allowed by the design of the instrument. Consequently, the internal buffer overflowed which resulting in an effectively different sampling rate than assumed. The mystery of the abnormal *dP/dt* was finally resolved.

After about ten days, Heiles and Backer returned to Berkeley. I stayed on because Backer had arranged a collaboration with Joseph Taylor (Princeton) who had a well-oiled analog filter-bank-based pulsar timing machine (for timing PSR 1913+16, the first binary pulsar) already at Arecibo. I quickly learnt the ropes from Taylor and started to time the millisecond pulsar. In late November Backer submitted the pulsar discovery paper to *Nature* (with 4C21.53 → PSR 1937+21). The manuscript was rapidly accepted and formally published on 16 December 1982 (Backer et al. 1982b). Apparently, *Nature* commissioned David Helfand (Columbia University) to write a commentary ("News & Views") on the discovery (Helfand, 1982). Helfand noted that A. Alpar, J. Shaham and M. Ruderman at Columbia University has surmised PSR 1937+21 was not a young object, but an old neutron star spun up by accretion from a companion (see Alpar et al. 1982). From India, Radhakrishnan & G. Srinivasan (1982) had independently reached the same conclusion. Indeed, the timing observations undertaken with Taylor's timing machine found *dP/dt* = $(1.25 \pm 0.25) \times 10^{-19}$ s/s (Backer et al., 1983). The implied slowing down timescale is about 0.2 billion years, somewhat shorter than anticipated in Helfand's commentary.

## 5      Epilogue

After a month's break I returned to Arecibo to undertake observations with my interferometer. In March 1983 I had a strong urge to "move on". Heiles agreed that I had done enough for my PhD thesis. I buttoned down, submitting my thesis in September 1983 (Kulkarni 1984a). I stayed on as a post-doctoral fellow, working with Backer, to design and build the "Fast Pulsar Timing Machine" (Kulkarni et al., 1984b).

Recall that 4C21.53 was a bright steep spectrum radio source identified in the 4C radio survey which showed interplanetary scintillation despite being located at low Galactic latitudes (see Readhead 2024 for a detailed account). Alan Purvis, a graduate student at Cambridge University, working with Anthony (Tony) Hewish, was undertaking a new IPS survey at 81.5 MHz (Purvis et al., 1987). During 1984-1985, working with Purvis our group undertook two programs to find more objects like 4C21.53 (steep spectrum, low Galactic latitude sources with promise of IPS scintillation). We conducted pulse and scintillation searches at Arecibo (Purvis et al. 1984, Heiles et al. 1984), but with no success.

I note that the primary output of my thesis consisting of three papers: Dickey et al. 1983, Kulkarni et al. 1985, Kulkarni et al. 1985b, which even by the standards of those days was quite thin. Nonetheless, Caltech offered me a prestigious Millikan Fellowship in Experimental Astrophysics. I suspect that this happy outcome was due to energetic lobbying by Readhead. In the summer of 1985, I got married and left for Caltech in Fall. At Caltech I began a dual program, one centered on optical and NIR interferometry and another, spurred by a paper by Hamilton et al. (1985), to explore the connection between low-mass X-ray binaries (LMXBs) and millisecond pulsars. The latter effort led to what I consider as one of my original papers on neutron stars (Kulkarni, 1986) and, working with my friend and colleague John Middleditch at Los Alamos National Laboratory (which hosted a Cray XMP supercomputer equipped with a vector processor), to the discovery of the first pulsar in a globular cluster (Lyne et al., 1987).

## 6    NOTES

1. Incidentally, my investment in planetary radar paid off later in my career: my student developed a method to determine the orientation of asteroid rotation based on detection of radar signals across the VLBA antennas (Busch, 2010).

**Acknowledgements:** This account was written up at the request of my dear friend and colleague, Dr. Miller Goss, National Radio Astronomy Observatory (NRAO) and Professor Wayne Orchiston, one of the Co-Editors of *JAHH*. It is meant to accompany two other articles: Demorest & Goss (2024) and Readhead (2024). I thank Paul Demorest (NRAO) for prodding me to complete the manuscript.